\documentclass[aps,twocolumn,nofootinbib,superscriptaddress,floatfix]{revtex4}

\usepackage{color,amsmath,amssymb,graphicx,latexsym,subfigure}
\usepackage{threeparttable}
\usepackage[colorlinks]{hyperref}
\usepackage{ulem,soul}
\newcommand{\gev}{\,{\rm GeV}}

\newcommand{\beq}{\begin{equation}}
\newcommand{\eeq}{\end{equation}}
\newcommand{\bea}{\begin{eqnarray}}
\newcommand{\eea}{\end{eqnarray}}

\newcommand{\gsim}{\lower.7ex\hbox{$\;\stackrel{\textstyle>}{\sim}\;$}}
\newcommand{\lsim}{\lower.7ex\hbox{$\;\stackrel{\textstyle<}{\sim}\;$}}

\newcommand{\be}{\begin{equation}}
\newcommand{\ee}{\end{equation}}
\newcommand{\ba}{\begin{eqnarray}}
\newcommand{\ea}{\end{eqnarray}}

\newcommand{\SgrA}{Sgr~A$^*$}

\begin{document}

\title{Constraining ultralight bosonic dark matter with Keck observations of S2's orbit and 
kinematics} 

\author{Guan-Wen Yuan}
\affiliation{Key Laboratory of Dark Matter and Space Astronomy, Purple Mountain Observatory, Chinese Academy of Sciences, Nanjing 210023, China}
\affiliation{School of Astronomy and Space Science, University of Science and Technology of China, Hefei 230026, China}

\author{Zhao-Qiang Shen}\email{zqshen@pmo.ac.cn}
\affiliation{Key Laboratory of Dark Matter and Space Astronomy, Purple Mountain Observatory, Chinese Academy of Sciences, Nanjing 210023, China}

\author{Yue-Lin Sming Tsai}\email{smingtsai@pmo.ac.cn}
\affiliation{Key Laboratory of Dark Matter and Space Astronomy, Purple Mountain Observatory, Chinese Academy of Sciences, Nanjing 210023, China}

\author{Qiang Yuan}
\affiliation{Key Laboratory of Dark Matter and Space Astronomy, Purple Mountain Observatory, Chinese Academy of Sciences, Nanjing 210023, China}
\affiliation{School of Astronomy and Space Science, University of Science and Technology of China, Hefei 230026, China}

\author{Yi-Zhong Fan}
\affiliation{Key Laboratory of Dark Matter and Space Astronomy, Purple Mountain Observatory, Chinese Academy of Sciences, Nanjing 210023, China}
\affiliation{School of Astronomy and Space Science, University of Science and Technology of China, Hefei 230026, China}

\begin{abstract}
Ultralight bosonic dark matter is expected to be able to form a cloud surrounding the supermassive 
black hole (SMBH) in the Galactic center. With increasing precision of the observations of the
stellar kinematics around the SMBH, tiny effects from such a dark matter cloud, including its
gravitational perturbation and the direct coupling with the ordinary matter may be detectable. 
In this work, we search for possible evidence of the scalar cloud using accurate orbital measurements of the S2 star around Sgr~A*. 
We solve the first order Post-Newtonian equation, considering simultaneously the extended mass distribution of the scalar cloud 
and the frequency shift induced by the additional coupling via Higgs portal or photon portal interaction. 
Furthermore, we also investigate the impact of an astrophysical power-law component from the gas and stellar remnants.
We find that the astrometric and spectroscopic data of the S2 star are well consistent with the scenario of a point-like mass of Sgr~A*. 
We thus derive upper limits of the coupling of the new interaction and the extended mass, with and without the contribution from the astrophysical component.
The limits of the Higgs/photon coupling and the extended mass of the scalar cloud are the most stringent
ones for the scalar mass window between $3.2\times 10^{-19}$~eV and $1.6\times 10^{-18}$~eV.  

\end{abstract}

\date{\today}
\maketitle
 
\section{Introduction}
One of the most attractive physics beyond the standard model (SM) is the nature of dark matter (DM), which is also one of the most puzzling topics in cosmology and astrophysics~\cite{Rubin:1980zd, Bertone:2004pz,Clowe:2006eq,Feng:2010gw}. 
Besides those gravitational DM evidence, we are still unclear about the particle nature of DM after decades of efforts in
DM direct detection~\cite{Liu:2017drf, Chen:2020gcl, PandaX-II:2021kai}, indirect detection~\cite{Fermi-LAT:2013sme, Daylan:2014rsa,Gaskins:2016cha,PPTA:2021uzb}, and collider experiments~\cite{ATLAS:2012ky,Boveia:2018yeb,Buchmueller:2017qhf}.  
The dynamics of the stars around the supermassive black hole (SMBH) can provide an interesting 
but independent way to probe the DM. 
Several well-motivated DM candidates could be observable near the SMBH, 
such as the weakly interacting massive particles~\cite{Jungman:1995df, Arcadi:2017kky, Roszkowski:2017nbc,Yuan:2021mzi}, 
and ultralight DM coherent or bound by strong gravitation~\cite{Schive:2014dra,Hui:2016ltb,Zhang:2016uiy,Chen:2019fsq,Hui:2021tkt,Sun:2021yra}.

In the last decades, with the development of interferometric imaging and 
high-precision narrow-angle astrometry in the infrared bands, 
the experimental uncertainties of the stellar kinematics in the Galactic center are significantly improved~\cite{Ghez:1998ph, Eckart:2002in,Ghez:2003rt, Gillessen:2008qv, Do:2014uca}.
Moreover, SMBHs, one kind of astrophysical extreme compact object, is found in most galaxies, and we understand their properties better and better recently, thanks to rapidly improving observations.
Two of the most well-known experiments are performed by the Keck Observatory~\cite{Ghez:2003qj, Do:2014uca, Do:2019txf, Hees:2020gda} and 
the Very Large Telescope (VLT)/GRAVITY~\cite{2017ApJ83730G, GRAVITY:2018ofz, GRAVITY:2019zin, GRAVITY:2020gka, GRAVITY:2021xju}. 
They have been observing the motion of short-period stars orbiting around the Galactic center, Sagittarius A$^{*}$ (\SgrA), for over 20 years. 
The observations of such extreme environments with strong gravity are good probes of the properties of the SMBH~\cite{Kormendy:1995er, Ferrarese:2004qr,Ghez:2008ms,Genzel:2010zy,Chen:2018yje}, the star and galaxy formation~\cite{Bahcall:1976aa,Quinlan:1994ed,Alexander:1999yz,Ferrarese:2000se,Alexander:2005jz}, the tests of general relativity or other gravity theories~\cite{Baker:2014zba,Will:2014kxa,Yan:2019hxx}, the nature of DM~\cite{Sadeghian:2013laa,Buckley:2017ijx, Lacroix:2018zmg, Bar:2019pnz, GRAVITY:2019tuf, Nampalliwar:2021tyz, Tsai:2021irw}, and so on.

Among various candidates, ultralight bosonic DM, likely to form a Bose-Einstein condensation, 
is an attractive cold DM candidate~\cite{Hu:2000ke, Amendola:2005ad, Schive:2014hza, Kouvaris:2019nzd, Sun:2019ico, Yuan:2020xui, Chen:2021lvo, Du:2022trq}. 
Within this framework, the possible interaction between the SM particles and bosonic DM will result in a frequency shift via either Higgs or photon portal interaction~\cite{Stadnik:2015kia,Kouvaris:2019nzd,Kouvaris:2021phj}. 
The electron mass or fine structure constant can hence receive an additional correction. 
Thus, we propose to search for such ultralight bosonic DM, by including both the perturbations on the stellar orbits from the extended DM mass and the velocities from the direct DM-SM couplings for the first time. In addition, an astrophysical power-law component from gas and stellar remnants is also considered as the other case.
In this paper, we only use the publicly available position and velocity data of the S2 star observed by the Keck Observatory~\cite{Do:2019txf}.
For the density profile of such a scalar field in the Galactic center, we adopt the model developed for the black hole-scalar field system~\cite{Bar:2019pnz, Nampalliwar:2021tyz}, which predicts the formation of a scalar cloud. 
We perform the Bayesian analysis using the Markov Chain Monte Carlo method (MCMC) to scan the DM and stellar orbital parameters. 
Finally, stringent upper limits for the Higgs/photon portal coupling and DM extended mass are calculated.

This paper is outlined as follows. 
In section \ref{superradiance}, we revisit the black hole-scalar field system, 
and calculate the density profile of the scalar field. 
In section \ref{interaction}, we discuss the Higgs portal or photon portal interaction between the scalar field and SM, which results in a frequency shift of hydrogen emission lines. 
In section \ref{data_analysis}, we provide the methodology of data analysis in orbital fitting, including the orbital model and statistical analysis framework. 
Then, we present our constraints on the Higgs coupling $\beta$, the photon coupling $g$ and the extended mass $M_{\rm ext}$ in section \ref{bounds}, and summarize this work in section \ref{conclusion}.

\section{The black hole-scalar field system}\label{superradiance}
The massive scalar field $\Phi$, described by the Klein-Gordon equation, 
can develop a condensed structure around the black hole through the superradiance mechanism
\cite{Detweiler:1980uk,Strafuss:2004qc,Dolan:2007mj, Sikivie:2009qn, Yoshino:2012kn,Witek:2012tr, Guth:2014hsa, Brito:2015oca,Kobayashi:2017jcf,Amorim:2019hwp, Zu:2020whs,Roy:2021uye}.
In this section, we discuss the solution of the Klein-Gordon equation in a Kerr metric, whose action is written as 
\begin{equation}\label{action}
S=\int d^{4} x \sqrt{-g}\left(\frac{R}{16 \pi}-\frac{1}{2} g^{\alpha \beta} \Phi^{*}_{, \alpha} 
\Phi_{, \beta}^{*}-\frac{\mu_s^{2}}{2} \Phi \Phi^{*}\right),
\end{equation}
where $R$ is the Ricci scalar, $g_{\alpha\beta}$ and $g$ are the metric tensor and its determinant, 
$\Phi(t, r, \theta, \varphi)$ is a complex scalar field with $\Phi_{, \alpha}=\partial \Phi /\partial x_\alpha$,
and $\mu_s$ is the mass of $\Phi$.
The scalar field around SMBH follows the Klein-Gordon equation
\begin{equation}
    \nabla_{\alpha} \nabla^{\alpha} \Phi= \mu_s^{2} \Phi
\end{equation}
and the Einstein equation
\begin{equation}
    G^{\alpha \beta}= 8 \pi T^{\alpha \beta},
\end{equation}
where $G^{\alpha\beta}\equiv R^{\alpha\beta}-\frac{1}{2}Rg^{\alpha\beta}$ is the Einstein tensor, 
$T^{\alpha \beta}$ is the energy-momentum tensor of scalar field.

Usually, the influence of the scalar field on the metric is not that large comparing to the SMBH, so we can adopt a fixed background metric for a stationary spin black hole, which is written in the standard Boyer-Lindquist coordinate 
\begin{equation}\label{kerr_metric}
\begin{aligned} 
d s^{2}=&-\left(1-\frac{2 M_{\bullet} r}{\Sigma}\right) d t^{2} 
-\frac{4 a M_{\bullet} r}{\Sigma} \sin ^{2} \theta d t d \varphi \\ 
&+\frac{\Sigma}{\Delta} d r^{2}+\Sigma d \theta^{2} \\ 
&+\left(r^{2}+a^{2}+\frac{2 M_{\bullet} a^{2} r}{\Sigma} \sin ^{2} \theta\right) 
\sin ^{2} \theta d \varphi^{2} ,
\end{aligned}
\end{equation}
with
\begin{equation}
\begin{aligned}
&\Sigma = r^2 + a^2 \sin^2\theta,\\
&\Delta = r^2 - 2 M_{\bullet}r + a^2 \equiv (r-r_{+})(r-r_{-}), 
\end{aligned}
\end{equation}
in which $r_{\pm}=M_{\bullet}\pm\sqrt{M^2_{\bullet}-a^2}$ with the dimensionless angular momentum $J=a M_{\bullet}$ and the spin parameter $a$.
The quantity $M_{\bullet}$ is the black hole mass. 
Under the Kerr metric, the solution of the scalar field can be described as
\begin{equation}\label{solution_form}
\Phi_{\ell m}(t, r, \theta, \varphi)= e^{-iwt +i m \varphi}S_{\ell m}(\theta)R_{\ell m}(r), 
\end{equation}
where $\ell, m$ correspond to the angular modes and $w$ is the frequency of the scalar field, including the real component $w_R$ and the imaginary component $w_I$. 

In addition, to describe the mass coupling of the black hole-scalar field system in the exponential term in Eq.~\eqref{solution_form}, we define a dimensionless quantity with physical clarity $\zeta$~\cite{Amorim:2019hwp},
\begin{equation}
\zeta=\frac{r_g}{\lambda_C}=\mu_s M_{\bullet} \sim \mathcal{O}(10^{-3}),
\end{equation}
where $r_g=M_\bullet$ in Planck units is the gravitational radius of the black hole, and  $\lambda_C =\mu_s^{-1}$ in Planck units is the Compton wavelength of the particle with mass $\mu_s$. 

We would like to note that the mass coupling $\zeta \ll 1$ in our case, 
because $\lambda_C$ is much larger than the gravitational radius of the black hole. 
We can obtain the time dependence of the quasi-bound-state frequencies $w = w_R + iw_I$, whose real and imaginary components are both positive and very small~\cite{Amorim:2019hwp}
\begin{equation}\label{frequency_components}
\left\{\begin{array}{ll}
w_{R} & \sim \mu_{s}-\mu_{s}\left(\frac{\zeta}{\ell+n+1}\right)^{2}, \\ 
w_{I} & \sim \mu_{s}\left(\frac{am}{M_{\bullet}}-2 \mu_{s} r_{+}\right) \frac{\zeta^{4 \ell+4}}{\sigma_{\ell}},
\end{array}\right.
\end{equation}
where spin parameter $a$ only appears in $w_{I}$, 
and $\sigma_{\ell}$ depends on the system~\cite{Brito:2015oca}.
The positive $w_{I}$ indicates that the scalar field profile will develop exponentially with time, 
characterized by time scale $w_I =1/ \tau_I$. 
Taking the fastest growing mode of Sgr~A* ($n=0$, $l=m=1$) as an example, 
its time scale is given by $\tau_I \sim \zeta^{-9}$.
On the other hand, $w_R= 1/ \tau_R$ describes the oscillation of the scalar field.  
With the limit $\zeta \ll 1$, the oscillation time is much smaller than 
the growth time scale, namely $\tau_R\sim \zeta^{-1}\ll \tau_I$. 
Thus, we can safely ignore the growth effect of the scalar field, 
instead of focusing on the oscillation effect at the level of observations.

We now turn our attention to the spheroidal harmonic function $S_{\ell m}(\theta)$ and the radial function $R_{\ell m}(r)$ in Eq.~\eqref{solution_form}, which can be written as~\cite{Detweiler:1980uk} 
\begin{equation}\begin{aligned} \label{spheroidal_harmonics}
\frac{1}{\sin \theta} \frac{d}{d \theta}\left(\sin \theta \frac{d S_{\ell m}}{d \theta}\right) &
-\left[a^2q^2 \cos ^{2} \theta +\frac{m^{2}}{\sin ^{2} \theta}\right] S_{\ell m} \\
&= A_{\ell m}S_{\ell m}, 
\end{aligned}\end{equation}
and
\begin{equation}\label{radial_function}
\begin{aligned} 
\frac{d}{d r}\left(\Delta \frac{d R_{\ell m}}{d r}\right)&+\left[\frac{K^{2}}{\Delta}-a^{2} w^{2}+ 2maw- \mu_s^{2} r^{2} \right] R_{\ell m} \\ 
&= A_{\ell m}R_{\ell m}.
\end{aligned}
\end{equation}
where we define $q\equiv\sqrt{\mu_s^2-w^2}$ and $K\equiv (r^2 +a^2)w -a m$, and $A_{\ell m}$ is the eigenvalue corresponding to the radial and angular parts of the Klein-Gordon equation. 
Again, as the scalar mass $\mu_s$ we detected is very small and the growth of dynamics of the scalar field is too small, the scalar field is effectively stationary over the time of observations.
As seen in Eq.~\eqref{frequency_components} that the Kerr spin parameter $a$ only appears in $w_I$ 
and not affects the scalar field oscillation, 
we can ignore the terms involving $a$ in the wave equation. 
Therefore, we can simply obtain the solution of Eq.~\eqref{spheroidal_harmonics} and
Eq.~\eqref{radial_function} as in Refs.~\cite{Detweiler:1980uk,Brito:2015oca,Amorim:2019hwp}. 
\begin{equation}\label{SR_solution}
\left\{\begin{array}{ll}
S_{\ell m}(\theta) & =P_{\ell}^{m}(\cos \theta), \\
R_{\ell n}(r) & =A_{\ell n} v^{\ell} \mathrm{e}^{-v / 2} L_{n}^{2 \ell+1}(v), \end{array}\right. 
\end{equation}
in which $n$ is an integer number for identifying the solution, $A_{\ell n}$ is a normalization constant, $P_{\ell}^m$ and $L_n^{2\ell +1}$ are 
Legendre polynomials and generalized Laguerre polynomials, respectively. 
We define 
\begin{equation}
v=\frac{2rM_{\bullet}\mu_s^2}{\ell+n+1},
\end{equation}
and $r$ in Planck units is the distance to the center of black hole in the radial coordinate. 

As long as a very small $\mu_s$ is applied, one can safely disregard the $w_I$ of the exponential term in Eq.~\eqref{solution_form} 
and the contribution from $a$ in the wave equation. 
Here, we just consider the specific mode ($n=0$, $l=m=1$), 
as this field mode grows more efficiently due to the superradiance mechanism.
These simplifications are also used in Refs.~\cite{Detweiler:1980uk, Amorim:2019hwp}.
Finally, the scalar field profile in this black hole-scalar field system can be written as
\begin{equation}\label{scalar_field3}
\Phi=C_{0} \mathrm{e}^{-\mathrm{i}\left(\bar{w}_{R} \bar{t}-\varphi\right)} \bar{r} \zeta^{2} 
\mathrm{e}^{-\bar{r}\zeta^{2}/2} \sin \theta .
\end{equation}
where the time-independent constant $C_0$ is related to the total mass of the scalar cloud. 

It is worth mentioning that we have used the normalized coordinates in units of $M_{\bullet}^{-1}$,
\begin{equation}
\begin{aligned}
\bar{w}_R &\propto \zeta -\zeta \left(\frac{\zeta}{\ell +n+1}\right)^2,\\
\bar{r}&=\frac{r}{M_{\bullet}} \text{ and }  \bar{t}=\frac{t}{M_{\bullet}}.
\end{aligned}
\end{equation}
In the limit of $\zeta \ll 1$, the energy-momentum tensor of the scalar field is dominant by the low energy limit.
Hence, the scalar field density distribution is well described by ~\cite{Amorim:2019hwp}
\begin{equation}\label{density}
\rho = \frac{1}{2}\mu_s^{2}|\Phi|^{2}=\frac{1}{2}\mu_s^{2}\left(C_{0}^{2} \zeta^{4} \mathrm{e}^{-\zeta^{2} \bar{r}} \bar{r}^{2} \sin ^{2} \theta\right).
\end{equation}

If the scalar fields is mainly described by the fastest growing mode, 
we obtain the total mass of the scalar cloud $M_{\rm cloud}$ by integrating full space. To describe the scalar cloud, 
we introduce dimensionless parameter $\kappa$ to replace the constant $C_0$, 
\begin{equation}\label{kappa}
\kappa = \frac{M_{\rm cloud}}{M_{\bullet}} =\frac{\int \rho r^{2} \sin \theta d r d \theta d \varphi}{M_{\bullet}}.
\end{equation}
The density distribution is varied with $\mu_s$, $\kappa$, and $r$.
And the total mass of the scalar cloud depends on $\mu_s$ and $\kappa$.

\begin{figure}[htbp]
\begin{center}
\includegraphics[width=0.98\linewidth]{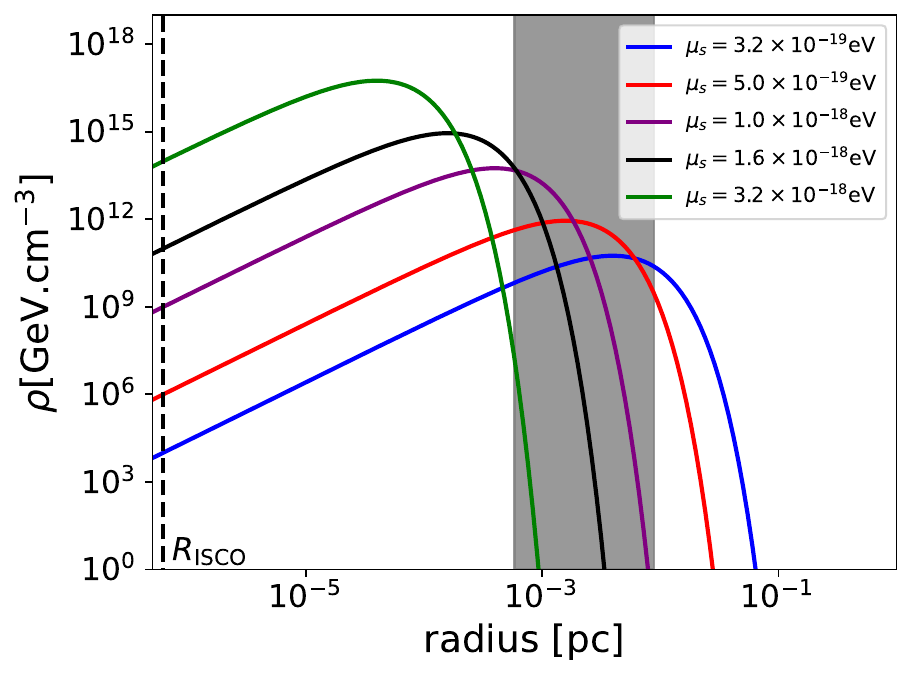}
\end{center}
\caption{The density distribution of scalar cloud around \SgrA. Gray region is the distance covered by the S2 orbit, and the scalar field mass $\mu_s = 3.2\times 10^{-19}, 5.0\times 10^{-19}, 1.0\times 10^{-18}, 1.6\times 10^{-18}, 3.2\times 10^{-18}$eV are also shown with blue, red, purple black, and green lines, respectively. The vertical black dotted line shows corresponds to the innermost stable circular orbit of \SgrA.
We fix the mass of the scalar cloud as $4 \times 10^{3} M_{\odot}$ (corresponding to $\kappa\sim 0.001$).}
\label{density_profile}
\end{figure}

Assuming the mass of the scalar cloud is $4 \times 10^{3} M_{\odot}$~\cite{GRAVITY:2020gka}, we display the DM density distribution of \SgrA  with different lines in Fig.~\ref{density_profile}.  
In addition, the innermost stable circular orbit of \SgrA is presented by the black dotted line and 
the radius covered by the S2 orbit is also shown with the shaded region. 
We can see that the density of the scalar field in the mass window $(3.2-16)\times 10^{-19}$~eV peaks between the pericenter and the apocenter of S2.

\section{Interacting with SM and Frequency Shift}\label{interaction}

The data used to measure the radial velocities of S2 are the K-Band (2.0 to 2.5 $\mu$m) spectra. The Brackett-$\gamma$ (Br$\gamma$) line, the dominant absorption feature, is among the strongest lines in the K-band, which is produced by the hydrogen transition~\cite{2017ApJ847120H}. 
The energy levels of the hydrogen atom follow
\begin{equation}
E_{n}=-\frac{\mu e^{4}}{2\left(4 \pi \epsilon_{0}\right)^{2} \hbar^{2}} \frac{1}{n^{2}} \simeq -\frac{m_e c^2}{2}\frac{\alpha^2}{n^2},
\end{equation}
where we use the definition of the fine structure $\alpha=\frac{e^2}{(4 \pi \epsilon_0) \hbar c}$ and approximate the reduced electron mass with $\mu=\frac{m_e m_p}{m_e + m_p}\simeq  m_e$. Thus the energy of the Br$\gamma$ line is
\begin{equation}\label{emission_line}
V_{mn}=E_m - E_n = -\frac{m_e c^2}{2}\left(\frac{1}{m^2} -\frac{1}{n^2}\right)\alpha^2 ,
\end{equation} 
in which $m, n$ are the different transition levels.

\subsection{Higgs portal model}
Considering the ultralight bosonic DM interacting with SM particles via the Higgs portal~\cite{Patt:2006fw, Cheung:2012xb, Cline:2013gha, Martins:2017yxk, Choi:2018axi, Bovy:2008gh, Barkana:2018lgd, Tsai:2021irw}, 
the additional Lagrangian to the SM can be written as     
\begin{equation}
\mathcal{L}_{\Phi H} = \beta |\Phi|^2 |H|^2,
\end{equation}
where $\beta$ is the dimensionless coupling between Higgs and bosonic DM.
This interaction generates a shift of the Higgs vacuum expectation value (VEV)~\cite{Kouvaris:2019nzd, Kouvaris:2021phj}
\begin{equation}
\hat{v}=v_{\mathrm{ew}} \sqrt{1-\frac{2 \beta}{m_{H}^{2}}\frac{\rho (r)}{2\mu_s^2}} \approx v_{\mathrm{ew}}\left(1-\frac{\beta}{m_{H}^{2}}\frac{\rho (r)}{2\mu_s^2}\right),
\end{equation}
where $v_{\rm ew}$ is the SM Higgs vacuum expectation value for $\beta=0$, $m_{H}$ is the Higgs mass, and $\rho (r)$ is the expectation value of the scalar cloud.
Therefore, the presence of the bosonic DM induces an effective change in the mass of SM particles. Taking the electron mass as an example, we have 
\begin{equation}
m_e \approx m_e^{\rm bare}\left(1-\frac{\beta}{m_H^2}\frac{\rho (r)}{2\mu_s^2}\right),
\label{eq:emass}
\end{equation}
where $m_e^{\rm bare}$ is the unperturbed electron mass.
Consequently, the frequency shift {$\delta f$} induced by DM can be written as
\begin{equation}
\frac{\delta f}{f} = \frac{\delta V_{mn}}{V_{mn}}.
\label{eq:fshift}
\end{equation}
Finally, based on Eq.~\eqref{emission_line}, \eqref{eq:fshift}, and \eqref{eq:emass}, 
we can rewrite the scalar cloud induced frequency shift for the Higgs portal DM model as
\begin{equation}\label{higgs_shift_f}
{\frac{\delta f}{f}(r) } = \left[\frac{\delta V_{mn}}{V_{mn}}(r)\right]_{\Phi H} \approx \frac{\delta m_e}{m_e} \approx \frac{\beta}{m_H^2} \frac{\rho (r)}{2\mu^2_s}.
\end{equation}
where $\delta f$ is a function of a position within the scalar cloud.
Here, we adopt the natural units for the conventions, namely $\rho$ in units of [$\gev^{4}$].

\subsection{Photon portal model}
Ultralight bosonic DM may also interact with the SM sector via photon, called photon portal model~\cite{Stadnik:2015kia}. 
The relevant interaction can be written as 
\begin{equation}
\mathcal{L}_{\Phi\gamma}=\frac{g}{4}|\Phi|^{2} F^{2},
\end{equation}
where $g$ is the dimensionless coupling between photon and bosonic DM, which can be either positive or negative. 
Therefore, the fine structure constant $\alpha$ has been altered to
\begin{equation}\label{eq:alpha}
\alpha = \alpha_0 \left(\frac{1}{1-gv_{\Phi}^2}\right) \approx \alpha_0 \left(1+g\frac{\rho}{2\mu_s^2}\right).
\end{equation}
Using Eq.~\eqref{emission_line}, \eqref{eq:fshift}, and \eqref{eq:alpha}, we can write the frequency shift for the  photon portal DM model
\begin{equation}\label{photon_shift_f}
\frac{\delta f}{f}(r) =\left[\frac{\delta V_{mn}}{V_{mn}}(r)\right]_{\Phi\gamma}
\approx \frac{2\delta \alpha}{\alpha} \approx 2g \frac{\rho (r)}{2 \mu_s^{2}}.
\end{equation}

\subsection{Frequency shift}
When the S2 star travels around Sgr A*, its gravitational potential varies dramatically. As a result, one can search for the ultralight bosonic DM evidence from the potential change, which is the strategy we utilized in Sec.~\ref{data_analysis}. 
The redshift from the hydrogen and helium absorption lines in the stellar spectrum can be observed as the S2 approaches the black hole. The Doppler effect of these lines can also be written as ~\cite{Rong:2017wzk, Berengut:2013dta, Hees:2017aal, Hees:2020gda, GRAVITY:2019zin},
\begin{equation}
\Delta f = f-f_0=\frac{v_e - v_o}{c}f_0=\frac{\Delta v}{c}f_0,
\end{equation}
and $\Delta v$ is the velocity difference. Thus, the velocity shifted by photon portal and Higgs portal can be described by ~\cite{Kouvaris:2021phj}
\begin{equation}\label{higgs_shift}
\Delta v_{\rm r, higgs}= \frac{\delta f}{f}(r) \approx \frac{\beta}{m^2_H}\frac{\rho (r)}{2\mu_s^2}.
\end{equation}
\begin{equation}\label{photon_shift}
\Delta v_{\rm r, photon}= \frac{\delta f}{f}(r) \approx 2g \frac{\rho (r)}{2 \mu_s^{2}},
\end{equation}
Again, we apply the natural units for the conventions.
S2 feels the variation of the scalar field density at different positions of the orbit, due to the changing of the gravitational potential.
Thus, we can determine the size of the frequency shift by measuring the S2 positions and velocities.

\section{Methodology}\label{data_analysis}

\subsection{Orbital model and initial conditions}
\label{sec:orbit_model}
\begin{figure}[htbp]
\begin{center}
\includegraphics[width=0.8\linewidth]{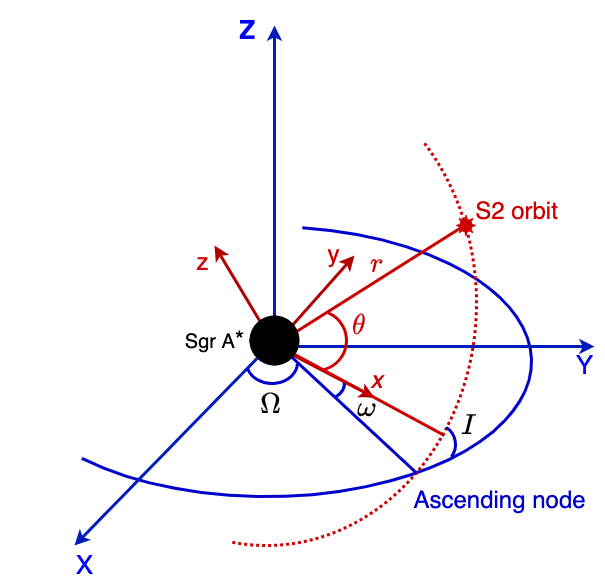}
\end{center}
\caption{Definition of the coordinate systems, whose axes originate from \SgrA, and projection of S2 orbit onto the plane of the sky.
The picture shows an illustration of the S2 star orbital parameters: $\theta$ is the azimuth angle of the spherical system of coordinates associated with the $x, y$, and $z$ Cartesian coordinates. For an elliptic motion in the $x-y$ plane, the two coordinates (red and blue) associated with each other via three Euler rotations: $I$ is the  angle of inclination between the S2 orbit and the observation plane, $\Omega$ is the angle of the ascending node and $\omega$ is the argument of pericenter. 
It is worth mentioning that the $Z$-axis of the coordinate system is defined by the vector pointing from the Galactic center to the solar system. 
}
\label{coordinate}
\end{figure}

The stellar orbit in the cloud of extended bosonic DM around the SMBH can be described with the following model parameters:
\begin{itemize}
    \item The mass of \SgrA ($M_{\bullet}$), which is the only physical parameter of the black hole.
    \item Six additional parameters about SMBH, including the distance $R_0$ between the Earth and \SgrA,
    the astronomical coordinates coinciding with the SMBH (Right Ascension $\alpha_{\rm BH}$, Declination $\delta_{\rm BH}$),
    the velocities to the coordinate origin $(v_{\alpha}, v_{\delta})$,
    and the line-of-sight velocity of SMBH relative to the Sun $v_z$.
    The current level of precision in radial velocity is $\sim 10~{\rm km/s}$, 
    so that acceleration terms are ignored~\cite{GRAVITY:2018ofz, Do:2019txf, Becerra-Vergara:2020xoj}. 
    \item Six parameters describe the stellar orbit in the coordinate system centering at the SMBH $(r_p, v_p, t_p, I, \Omega, \omega)$. 
    The radial distance and the tangential velocity are $r_{\rm p}, v_{\rm p}$ at the epoch of the closest approach $t_p$. 
    The angle of inclination, the position angle of the ascending node, and the argument of pericenter is denoted by $I, \Omega$, and $\omega$, respectively, as presented in Fig.~\ref{coordinate}.
    These parameters serve as the initial condition for stellar motion in the gravitational potential essentially.
    \item Two parameters about the bosonic DM: the mass ratio of the extended scalar cloud to the \SgrA mass $\kappa$, defined in Eq.~\eqref{kappa}, and the dimensionless coupling to the Higgs or photon ($\beta$ or $g$), also defined in Eq.~\eqref{higgs_shift}, \eqref{photon_shift}.
\end{itemize}
We have $1+6+6+2=15$ free parameters in total. 
In this work, we do not consider the spin and quadrupole of \SgrA, 
because these parameters only arise at 
the 1.5 and second order post-Newtonian (1.5PN and 2PN) 
approximation (proportional to $c^{-3}$ and $c^{-4}$) respectively~\cite{Will:2016pgm} 
and they are below the precision of current observation sensitivities. 
Hence, we only turn to the first order post-Newtonian (1PN) approximation.

The stellar orbit can be obtained by solving the following 1PN equation of motion~\cite{Do:2019txf}
\begin{equation}
\label{pN_equation}
\frac{{\rm d}^{2} \boldsymbol{r}}{{\rm d} t^{2}}=-\frac{G M}{r^{3}} \boldsymbol{r}+\frac{G M}{c^{2} r^{3}}\left(\frac{4G M}{r}-v^{2}\right) \boldsymbol{r}+\frac{4G M}{c^{2} r^{3}} (\boldsymbol{r} \cdot \boldsymbol{v})\boldsymbol{v},
\end{equation}
where $M\equiv M(r)$ is the enclosed mass, $\boldsymbol{r}(t)\equiv (X(t), Y(t), Z(t))$ and $\boldsymbol{v}(t)\equiv \boldsymbol{\dot{r}}(t)=(V_X(t), V_Y(t), V_Z(t))$ are the position and velocity of S2 with the origin located at the BH. 
As defined in Fig.~\ref{coordinate}, the coordinate system is centered on the \SgrA and S2 orbit is projected onto the plane of the sky. 
The direction of the $Z$-axis is points from the Galactic Center to the Solar System, and the $X-Y$ plane is parallel to the plane of the sky for the $X$-axis pointing West and the $Y$-axis pointing North.

We consider two cases for the mass model in Eq.~\eqref{pN_equation}.
In the first case (Case I), we assume that there are only two mass components within the S2's orbit, i.e. the SMBH and the scalar cloud:
\begin{equation}
\label{extmass}
M(r)=M_{\bullet} + 2\pi \int_{r_{\rm ISCO}}^{r} r'^{2} {\rm d}r' \int_0^\pi \sin\theta'\,{\rm d}\theta'\, \rho(r',\theta'),
\end{equation}
where $r_{\rm ISCO}$ is the radius of the innermost stable circular orbit, $\rho$ is the density of extended mass in Eq.~\eqref{density} which is parameterized with the mass ratio $\kappa$.
We also consider the other case (Case II) that an additional mass component from unresolved stars and gas exists in the very inner vicinity of the SMBH~\cite{Alexander:2017rvg,GRAVITY:2020gka, GRAVITY:2021xju}. Then the mass model is
\begin{equation}
\label{extmass2}
M(r)=M_{\bullet} + A r^{1.6} + 2\pi \int_{r_{\rm ISCO}}^{r} r'^{2} {\rm d}r' \int_0^\pi \sin\theta'\,{\rm d}\theta'\, \rho(r',\theta'),
\end{equation}
where a power-law stellar component $Ar^{1.6}$ with $A\simeq 2.516\times 10^{6} M_{\odot}$ as obtained by extrapolating the fitting results from large radii is added.

The stellar orbit in Eq.~\eqref{pN_equation} is a time-dependent second-order differential equation.
It can be solved with the {\tt DOP853} method~\citep{DOP853} implemented in {\tt SciPy}~\citep{scipy2020} given the initial condition of the star.
The initial state is defined at the pericenter epoch $t_p$ and the six phase-space coordinates can be transformed from the parameters $(r_p, v_p, I, \Omega, \omega)$ with the Euler rotations:
\begin{equation}\label{initial_coordinate}
\begin{aligned} 
X(t_p)   &= -r_p \sin\omega \cos I \sin\Omega + r_p \cos\omega \cos\Omega, \\ 
Y(t_p)   &=  r_p \sin\omega \cos I \cos\Omega + r_p \cos\omega \sin\Omega, \\ 
Z(t_p)   &= -r_p \sin\omega \sin I, \\
V_X(t_p) &= -v_p \cos\omega \cos I \sin\Omega - v_p \sin\omega \cos\Omega, \\
V_Y(t_p) &=  v_p \cos\omega \cos I \cos\Omega - v_p \sin\omega \sin\Omega, \\
V_Z(t_p) &= -v_p \cos\omega \sin I.
\end{aligned}
\end{equation}

The Romer time delay is a time modulation of a signal caused by the propagation of light through the S2 orbit in the $Z$-direction.
This effect is important for S2 in the Keck data ($-0.5$ days at pericenter and $7.5$ days at apocenter)~\citep{Do:2019txf}. 
Therefore, we account for the Romer time delay in our analysis.
The time delay between the emission point and the observer on the Earth can be described by 
\begin{equation}\label{Romer_delay}
t_{\rm em}=t_{\rm obs}+\frac{Z(t_{\rm obs})}{c}.
\end{equation}
Please note that we have different sign from that of~\citep{Do:2019txf}, since the $Z$ axis here is pointing from the Galaxy center to the Sun.
The Romer delay in Eq.~\eqref{Romer_delay} can be solved in an iterative way~\citep{Hees2014}.  
In our computation, we only perform one iteration, because the time correction obtained in the second iteration is already very small ($\lesssim 20$ minutes).
On the other hand, the Shapiro time delay and gravitational light deflection of the SMBH are ignored in our work, 
because they can only yield less than $\sim 5$ minutes and smaller than recent observational uncertainty, 
see Ref.~\cite{Do:2019txf} in more detail.

Once we set up Eq.~\eqref{pN_equation} with the initial conditions, 
the phase-space coordinate of S2 at any time $t$ can be predicted.
To compare with the observations, we convert the coordinate to the
right ascension (\texttt{R.A.}) $\alpha$, declination (\texttt{Dec.}) $\delta$ and radial velocity (RV) $v_r$ with
\begin{equation}\label{RA_Dec_RV_model}
\begin{aligned} 
\alpha_{*}(t_{\rm obs})&=\frac{Y(t_{\rm em})}{R_0}+\alpha_{\mathrm{BH}}+v_{\alpha}\cdot(t_{\rm em}-t_{\rm J2000}), \\
\delta_{*}(t_{\rm obs})&=\frac{X(t_{em})}{R_0}+\delta_{\mathrm{BH}}+ v_{\delta}\cdot(t_{\rm em}-t_{\rm J2000}),\\
v_r\left(t_{\mathrm{obs}}\right)&=
    V_{Z}\left(t_{\mathrm{em}}\right)+v_{r0}+
    \left[\frac{V_Z^{2}\left(t_{\mathrm{em}}\right)}{2 c}+\frac{G M}{c\,r(t_{\rm em})}\right]+\Delta V_r,
\end{aligned}
\end{equation}
where $t_{\rm J2000}$ is a reference epoch J2000.0, the four parameters $\alpha_{\rm BH}, \delta_{\rm BH}, v_{\alpha}$, and $v_{\delta}$ are the offset and linear drift relative to the reference frame center, and $v_{r0}$ is a constant velocity offset in the measurement of radial velocities.
The third term of radial velocity is the correction of relativity (the transverse Doppler shift and the gravitational redshift), which has been detected in~\cite{GRAVITY:2018ofz,Do:2019txf,GRAVITY:2020gka}.
The last term $\Delta V_r$ is the effect of frequency shift induced by the scalar cloud which is the main difference from the usual DM halo.
This frequency shift can be observed through the radial velocities.
It is related to the coupling constant ($\beta$ for Higgs portal, $g$ for photon portal) and the scalar field density, which has been discussed in Eq.~\eqref{higgs_shift} and Eq.~\eqref{photon_shift}.

\subsection{Statistical analysis}

\begin{figure*}[htbp] 
\centering 
\includegraphics[width=0.48\textwidth]{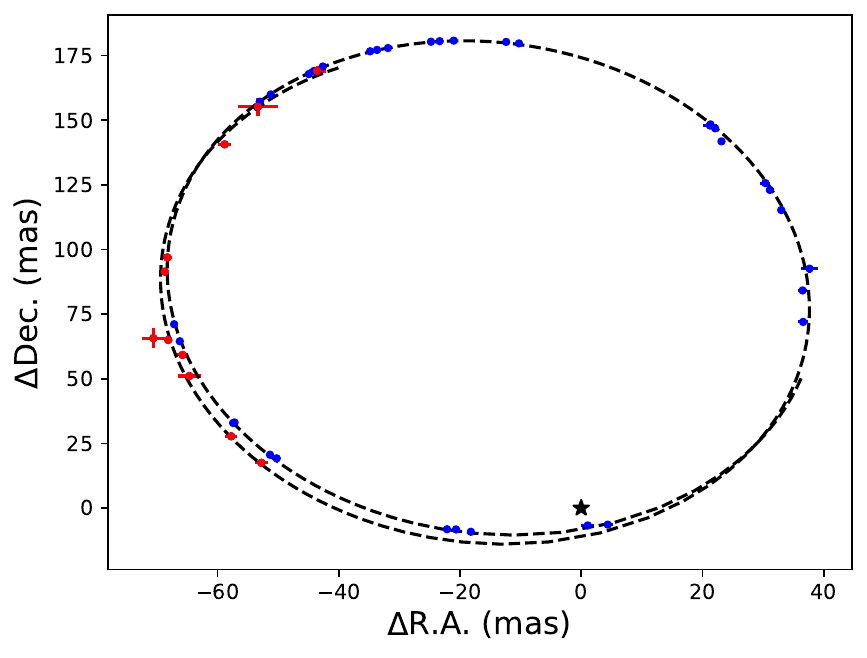} 
\includegraphics[width=0.48\textwidth]{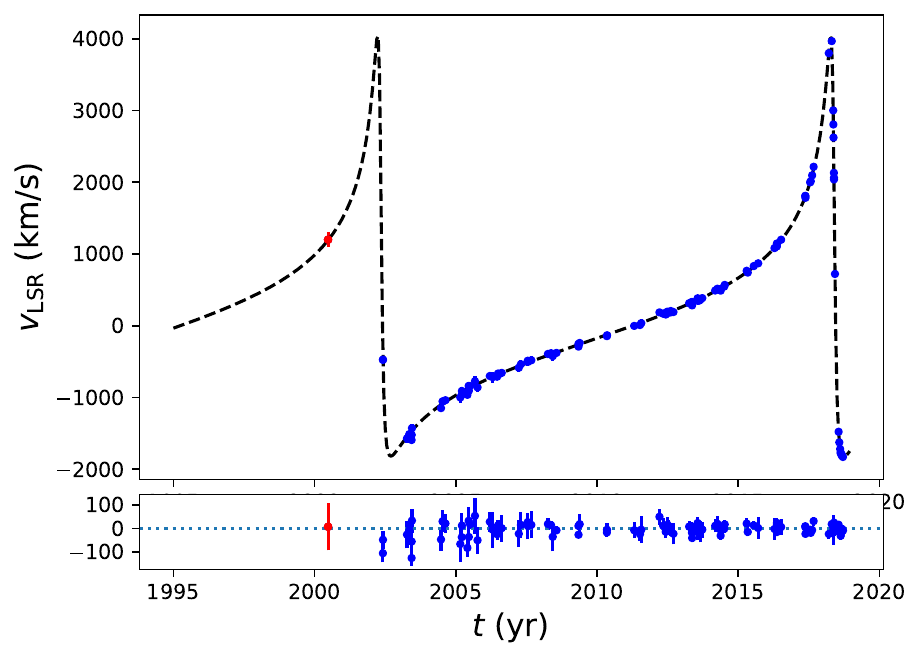}
\includegraphics[width=0.48\textwidth]{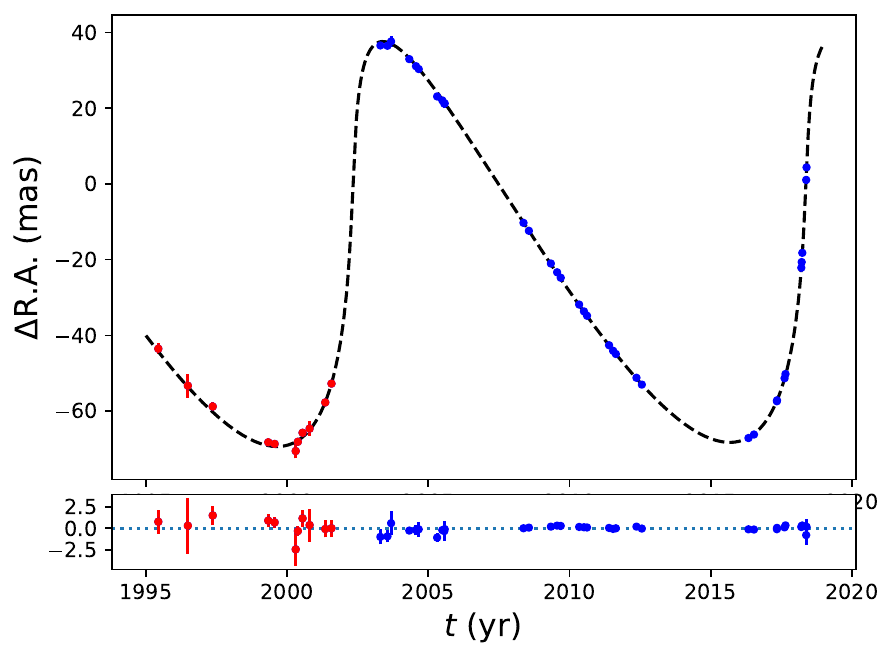} 
\includegraphics[width=0.48\textwidth]{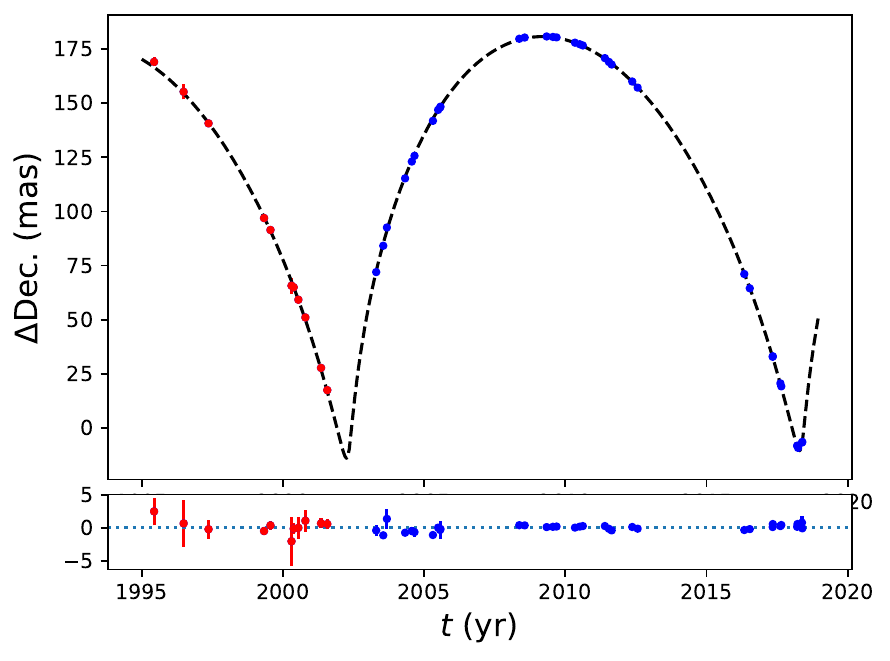}
\caption{The theoretical and observed orbit of S2 star around \SgrA. The theoretical fitting models are calculated by solving the equation of motion Eq.(\ref{pN_equation}). The upper left panel presents the 45 astrometric measurements of the S2 star orbit around \SgrA, overlaid with our best-fitting projected orbit in the plane of the sky. 
The black star is the place of \SgrA, and the axes of \texttt{R.A} and \texttt{Dec.} correspond to offset in right ascension and declination, respectively.
The upper right panel shows RV measurements and the best-fitting RV model using 115 RV measurements from 2000-2018. Residuals from the best-fitting RV model and error bars indicate $1\sigma$ uncertainties. The two panels at the bottom show the \texttt{R.A} and \texttt{Dec.} as a function of time and their respective residuals. 
Here, we use the observational data reported by Keck Observatory in~\cite{Do:2019txf}, and the best-fit parameters of the Higgs portal given in the third column of Table.~\ref{table1}. The corner of parameters has also presented in Fig.(\ref{contour}) in our Appendix.} 
\label{orbit_fitting}
\end{figure*}

In this work, we include 45 astrometric measurements (1995-2018) and 115 radial velocity measurements (2000-2018) from the Keck observations, which are publicly available in~\cite{Do:2019txf}. The total likelihood can therefore be separated into two parts
\begin{equation}
    {\mathcal L}_{\rm tot} = {\mathcal L}_{\rm astro} \times {\mathcal L}_{\rm RV}.
    \label{chi2}
\end{equation}

For the astrometric part, the correlations between nearby astrometric positions are considered with the following likelihood~\citep{Do:2019txf}
\begin{equation}
\begin{aligned}
{\mathcal L}_{\rm astro}\propto & (\left|{\boldsymbol\Sigma}_\alpha\right| \left|{\boldsymbol\Sigma}_\delta\right|)^{-\frac{1}{2}}\\
&\times \exp \left[-\frac{1}{2} \left(\Delta {\boldsymbol \alpha}^T {\boldsymbol\Sigma}_\alpha^{-1} \Delta {\boldsymbol \alpha}+\Delta {\boldsymbol \delta}^T {\boldsymbol\Sigma}_\delta^{-1} \Delta {\boldsymbol \delta} \right) \right],
\end{aligned}
\end{equation}
where $\Delta {\boldsymbol \alpha} \equiv \{\alpha_i - \mu_{\alpha}(t_{i})\}$ and $\Delta {\boldsymbol \delta} \equiv \{\delta_i - \mu_{\delta}(t_{i})\}$ are the vector of the differences between the observed and predicted \texttt{R.A.} and \texttt{Dec.}, respectively.
The covariance matrices for two positions
are $[{\boldsymbol\Sigma}_\alpha]_{ij} \equiv \sigma_{\alpha_{i}} \sigma_{\alpha_{j}} \boldsymbol{\rho}_{i j}$ and $[{\boldsymbol\Sigma}_\delta]_{ij} \equiv \sigma_{\delta_{i}} \sigma_{\delta_{j}} \boldsymbol{\rho}_{i j}$, 
where $\sigma_{\alpha}$ and $\sigma_{\delta}$ are the uncertainties of \texttt{R.A.} and \texttt{Dec.}, respectively.
The correlation matrix is given by $[\boldsymbol{\rho}]_{i j}=(1-p) \delta_{i j}+p \exp \left[-d_{i j}/\Lambda\right]$, 
where $d_{ij}$ is the angular distance between the point $i$ and $j$.
The correlation length scale $\Lambda$ and the mixing parameter $p$ will be fitted simultaneously with the model parameters.

For the radial velocities, the likelihood is simply obtained as 
\begin{equation}
    -2 \ln {\mathcal L}_{\rm RV} = \chi^2_{\rm RV} = \sum_j \left( \frac{v_{r,j}-\mu_{v_r}(t_j)}{\sigma_{v_r,j}} \right)^2,
\end{equation}
where $v_{r,j}$ and $\mu_{v_r}(t_j)$ are observed and predicted radial velocity observed at $t_j$, 
and $\sigma_{v_r}$ is its uncertainty.

A Bayesian analysis is performed with the Markov Chain Monte Carlo (MCMC) sampling tool \texttt{emcee}~\cite{emcee2013}.
Priors of the model parameters are given in Table.~\ref{table1}. We adopt Gaussian 
priors for the SMBH mass $M_{\bullet}$ and the Galactocentric distance of the Sun $R_0$.
The former is from the shadow of the SMBH imaged by the Event Horizion Telescope
(EHT)~\cite{EventHorizonTelescope:2022xnr}, and the latter is from the very long baseline interferometry (VLBI) parallaxes 
and proper motion of molecular masers in the spiral arms~\cite{2019ApJ885131R} which was
also used in determining the EHT measurement of $M_{\bullet}$. Log-uniform priors are 
assumed for the coupling constants ($\beta$ and $g$) and mass ratio $\kappa$ when scanning several orders of magnitude range, and uniform priors are assumed for the rest parameters.

In Fig.~\ref{orbit_fitting}, we present the orbit prediction based on the best-fitting Higgs portal model obtained from the MCMC scan.
The fitting parameters are shown in Table.~\ref{table1}, for a fixed scalar mass of $\mu_s = 10^{-18}$~eV.
The upper left panel is astrometric measurement of the S2 star's orbit around \SgrA (black star). 
The best-fitting orbit in the plane of the sky is presented as the dashed line. S2 moves counterclockwise in this projection. 
The origin of the coordinate system coincides with the location of \SgrA, as described in Fig.~\ref{coordinate}.
The upper right panel shows the best-fitting RV model, compared with the RV measurements from 2000 to 2018. The offsets of \texttt{R.A.} and \texttt{Dec.} in lower panels are relative to the position of \SgrA. 
The bottom sub-panels show the residuals between data and the model.

\begin{table*}
      {\small
      \begin{tabular}{lccccc}
            \hline
            \noalign{\smallskip}
             Parameter & Prior  &\multicolumn{2}{c}{Case I}&\multicolumn{2}{c}{Case II} \\
             &  & Best Fit  & Posterior mean $\pm 1\sigma$ & Best Fit  & Posterior mean $\pm 1\sigma$\\
            \noalign{\smallskip}
            \hline
            \noalign{\smallskip}
           $M_\bullet$ ($10^6 M_\odot$) & $ 4.0 \pm 1.1$  & 3.91 & $3.92 \pm 0.05$ & 3.90 & $3.92 \pm 0.05$  \\
           $R_0$ (kpc) & $ 8.15\pm 0.15 $ & 7.90 & $7.90 \pm 0.05$ & 7.89 & $7.91 \pm 0.05$ \\
           $\alpha_{\rm BH}$ (mas)& $(-10, 10)$  & $-1.38$ & $-1.27^{+0.35}_{-0.36}$ & $-1.28$ & $-1.22 \pm 0.33$  \\
           $\delta_{\rm BH}$ (mas) & $(-10, 10)$  & $-0.75$ & $0.79^{+0.38}_{-0.37}$ & $-0.87$ & $-0.75 \pm 0.37$ \\
           $v_{\alpha}$ ($\rm mas\,yr^{-1}$) & $(-500, 500)$ & 94 &  $87^{+19}_{-18}$ & 86 &  $85 \pm 18$ \\
           $v_{\delta}$ ($\rm mas\,yr^{-1}$) & $(-500, 500)$ & 220 & $221 \pm 20$ & 231 & $218 \pm 19$ \\
           $v_{r0}$ ($\rm km\,s^{-1}$) & $(-100, 100)$ & $-6.3$ & $-8.3 \pm 2.7$ & $-8.1$ & $-8.7 \pm 2.8 $ \\
           $r_p$ $(10^{-3}~\rm pc)$ & (0.01, 1) & 0.554 & $0.554 \pm 0.004$ & 0.552 & $0.554 \pm 0.003$ \\
           $v_p$ ($\rm km\,s^{-1}$) & $(-10^5, 10^5)$ & 7559  & $7571^{+28}_{-29}$ & 7559  & $7573^{+27}_{-26}$\\
           $t_p$ (yr) & (2010, 2030) & 2018.3818 & $2018.3818 \pm 0.0004$ & 2018.3819 & $2018.3818 \pm 0.0004$ \\
           $I$ ($^\circ$) & (0, 360) & 133.80 & $133.78 \pm 0.20$ & 133.70 & $133.80 \pm 0.18$ \\
           $\omega$ ($^\circ$) & (0, 360) & 66.70 & $66.66 \pm 0.12$ & 66.65 & $66.66 \pm 0.12$ \\
           $\Omega$ ($^\circ$)& (0, 360) & 227.83 & $227.77 \pm 0.16$ & 227.73 & $227.79 \pm 0.17$ \\
           $\rm offset$ ($\rm km\,s^{-1}$) & $(-300, 300)$ & 77 & $83^{+20}_{-21}$ & 81 & $83^{+19}_{-20}$ \\
           $p$ & (0.01, 0.99) & 0.70 & $0.54 \pm 0.13$ & 0.62 & $0.54^{+0.14}_{-0.13} $ \\
           $\Lambda$ (mas) & (1, 100) & 13 & $32^{+20}_{-17}$ & 16 & $31^{+19}_{-17}$ \\
           $\rm log_{10}\kappa$ & $(-6, -1)$ & $-4.45$ & $-4.79^{+0.96}_{-0.90}$ & $-5.51$ & $-4.82^{+0.95}_{-0.89}$  \\
           $\rm log_{10} \beta$ & $(-30, -20)$ & $-24.89$ & $-27.0^{+2.1}_{-2.2}$ & $-24.88$ & $-26.81^{+2.11}_{-2.20}$\\
           $\mu_s$ (eV)& --- & $10^{-18}$ & --- & $10^{-18}$ & --- \\
           $-2\ln {\mathcal L}_{\rm tot}$ & --- &  $-14.77$ & --- &  $-14.21$ & ---\\
           \noalign{\smallskip}
           \hline
           \hline
           \noalign{\smallskip} \noalign{\smallskip}
         \end{tabular} 
         }
\caption[]{\label{table1}
Priors and fitting results of all parameters in our work when fixing $\mu_s=10^{-18}$~eV.
Gaussian priors are assumed for the SMBH mass $M_{\bullet}$ and the Galacoocentric distance of the Sun $R_0$, 
log-uniform priors are assumed for coupling constants ($\beta$ and $g$) and mass ratio $\kappa$, 
and uniform priors are assumed for the rest parameters. 
For Case II, a power-law stellar component is added compared with Case I~\cite{GRAVITY:2020gka, GRAVITY:2021xju}. 
It should be emphasized that the mass of the SMBH is estimated from the shadow imaged by the EHT (we replace $4.0^{+1.1}_{-0.6}\times 10^6 M_{\odot}$ with $(4.0\pm 1.1) \times 10^6 M_{\odot}$ for
simplicity and being conservative)~\cite{EventHorizonTelescope:2022xnr}, 
and the distance is measured by the VLBI parallaxes and proper motions of molecular masers in the spiral arms~\cite{2019ApJ885131R}.} 

\end{table*}

\section{Constraints on Higgs and Photon portal Models}\label{bounds}
When the S2 star travels to different places in orbit, it will experience different scalar field density $\rho$ and 
the orbit can be altered by the extended mass and the frequency shift. 
During S2 orbital fitting, we not only consider the extended mass induced by the scalar field, but also the interaction between the ultralight bosonic DM and the SM via Higgs or photon portal, then setting strong limits to the extended mass ($M_{\rm ext}$) and coupling constants ($\beta$ and $g$).

\begin{figure}[htbp] 
\centering 
\includegraphics[width=0.5\textwidth]{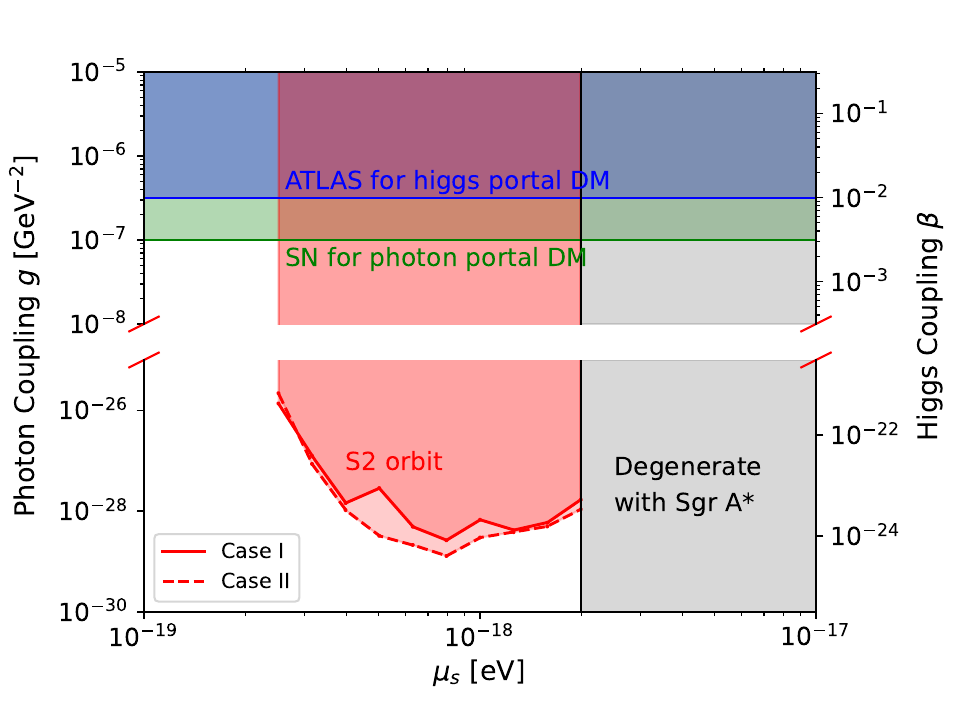} 
\caption{ The 95\% credible exclusion upper limits of the Higgs portal and photon portal coupling constants, without (solid) and with (dashed) the additional power-law astrophysical component.
The left and right $y$-axis are for the photon portal coupling $g$ and the Higgs portal coupling $\beta$ for different scalar field mass $\mu_s$.
The green and blue lines present the constraints from the supernova energy loss~\cite{Stadnik:2015kia} for the photon portal coupling and the ATLAS for the Higgs portal coupling~\cite{CMS:2018yfx}, respectively.}
\label{couple_limit}
\end{figure}

We are interested in the frequency shifts of atoms due to a bosonic DM surrounding the SMBH. 
For each DM mass $\mu_s$ between $10^{-19}$ eV and $10^{-17}$ eV, we scan over the total 15 model parameters as described in Sec.~\ref{sec:orbit_model}. 
We vary the Higgs coupling constant with a range of $10^{-30}\leq\beta\leq 10^{-20}$ for the Higgs portal DM, but the photon coupling constant is varied with $10^{-35} \rm{GeV}^2 \leq g\leq 10^{-25} \rm{GeV}^2$ for the photon portal DM. 
With these priors, the extended mass and couplings ($\beta$ and $g$) can simply escape from the current upper limits~\cite{Stadnik:2015kia, CMS:2018yfx}.

Using S2 data observed by Keck Observatory, we forecast the $95\%$ upper limits of Higgs portal and photon portal coupling in Fig.~\ref{couple_limit}.
The left and right $y$-axes represent the $\beta$ and $g$ scales, respectively. The ATLAS excludes the blue regions of Higgs coupling~\cite{CMS:2018yfx}, while the Supernova energy loss arguments exclude the green regions of photon coupling~\cite{Stadnik:2015kia}.
By assuming thermal equilibrium, the current photon coupling restrictions from the Big Bang Nucleosynthesis are more stringent than the limits derived in this work~\cite{Stadnik:2015kia}.

The bosonic DM density in the mass window of $(3.2-16.0)\times 10^{-19}$~eV is sensitive to the S2 orbit as seen in Fig.~\ref{density_profile}. 
If $\mu_s$ is too large, the radius of the scalar cloud is too short. 
The extended mass of DM can degenerate with SMBH mass $M_{\bullet}$ (gray region) because the uncertainties of radial velocity can overwhelm the smallish changes caused by the frequency shift. 
If DM is too light, the radius of the scalar cloud is too long. 
The extended mass is small so that the predicted frequency shift is negligible too. 
Thus, when the peak value of the scalar cloud is placed outside the S2 orbit, the constraints obtained in this work become weaker, 
see the behavior of the upper limit at $\mu_s<4\times 10^{19}$~eV region.

\begin{figure}[htbp] 
\centering 
\includegraphics[width=0.48\textwidth]{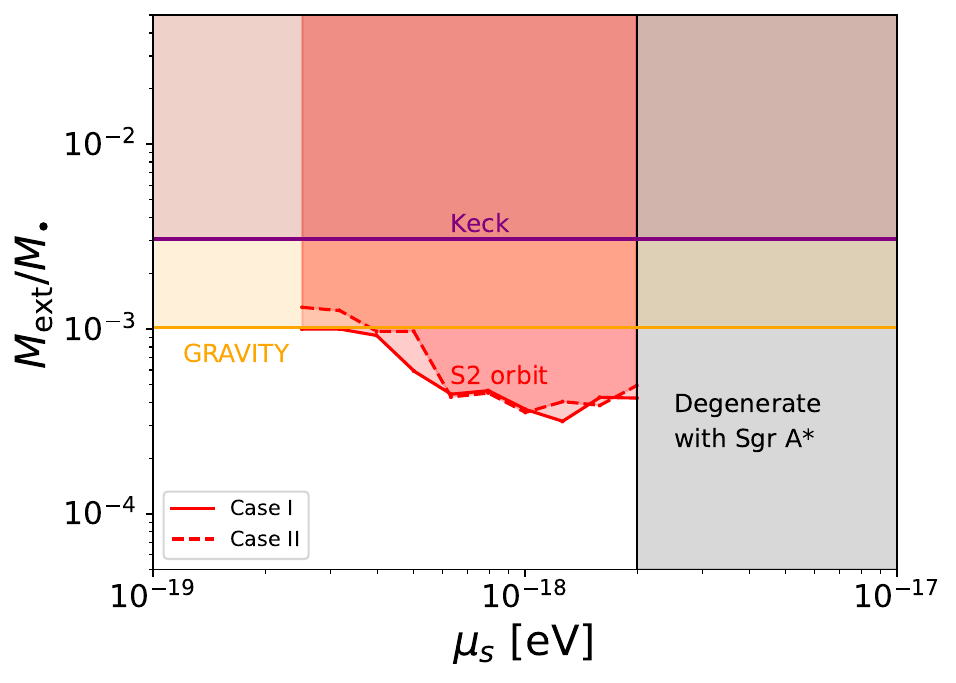} 
\caption{The 95\% credible upper limits of the extended mass from the scalar DM enclosed within the S2's orbit for different scalar field mass $\mu_s$, without (solid) and with (dashed) the additional power-law astrophysical component.
The Keck~\cite{Heissel:2021pcw} and GRAVITY~\cite{GRAVITY:2020gka} excluded regions on the extended mass are larger than $\sim 10^{4}~M_{\odot}$ and $\sim 4000~M_{\odot}$ within the radius of S2 orbit, as shown by purple and orange shaded regions.} 
\label{ratio_limit}
\end{figure}

In Fig.~\ref{ratio_limit}, we project the same upper limits in Fig.~\ref{couple_limit} to ($\mu_s$, $M_{\rm ext}$) plane with 
\begin{equation}
\label{extmass3}
M_{\rm ext}= 2\pi\int_{r_{\rm ISCO}}^{r_0} r'^{2} {\rm d}r' \int_0^\pi \sin\theta'\,{\rm d}\theta'\, \rho(r',\theta'),
\end{equation}
where $r_0$ is the apocenter of the S2 elliptical orbit, and $\kappa$ has defined in Eq.\eqref{kappa}. If the scalar cloud is within the range of the S2 orbit, the value of $\kappa$ is the same as $M_{\rm ext}/M_{\bullet}$. Otherwise, the constraints on $M_{\rm ext}/M_{\bullet}$ is weaker than $\kappa$, which can also be checked in Fig.\ref{density_profile}.
Clearly, the upper limits from S2 with 23 years of data are stronger than the result obtained by Keck~\cite{Heissel:2021pcw} and GRAVITY~\cite{GRAVITY:2020gka}. 
Comparing the constraints in Case I and Case II, we can see that the astrophysical power-law component ($\sim 1200~M_{\bullet}$ enclosed in S2's orbit~\cite{Alexander:2017rvg,GRAVITY:2020gka}) only has a relatively small impact on the parameters of the DM model. 
As expected, the constraints on the ultralight bosonic DM model parameters become more stringent when adding the astrophysical component. It is worth mentioning that our target DM component has induced a frequency shift by coupling with the SM particles and is thus distinguishable from the astrophysical component.

\section{Conclusion and discussion}\label{conclusion}
In this work, we propose a unique way to constrain the ultralight bosonic DM mass and its interaction 
with the SM via Higgs or photon portal. 
We consider a DM-induced frequency shift surrounding the SMBH in the center of our galaxy. 
Using the first order post-Newtonian approximation, we construct the stellar orbit equation including the DM-induced frequency shift, 
which also alters the radial velocity of S2.  
With the Keck S2 orbital data from the year 1995 to 2018, we then perform a Bayesian MCMC scan to estimate the 95\% credible upper limits of DM-Higgs interaction and DM-photon interaction. 
Similarly, upper limits of the extended mass of scalar cloud to DM particle masses $\mu_s$ are also computed.  
Because of S2 orbital size, only DM mass between $3.2\times 10^{-19}$~eV and $1.6\times 10^{-18}$~eV can be constrained by the Keck data. 
By marginalizing the nuisance parameters, our work shows that the DM-Higgs coupling 95\% upper limit ($\beta<\mathcal{O}(10^{-24})$ at $\mu_s\simeq 10^{-18}$~eV) is much stronger than 
the ATLAS limit. For the photon portal DM, the DM-photon coupling is also suppressed to be 
$g <\mathcal{O}(10^{-28})~\gev^{-2}$ at $\mu_s\simeq 10^{-18}$~eV which is also several orders better than 
the limit obtained by supernova energy loss arguments. 
The Keck data can also set an upper limit of the extended mass of scalar cloud, and it is about $3\times 10^{-4} M_{\bullet}$, which statistically agrees with the scenario of point-like mass of \SgrA. 
With ultralight DM-induced cloud involved in the fitting, we improve the extended mass upper limit from the latest GRAVITY results.
 
It is promising that there are lots of stellar remnants and gas around SMBH.  
By applying an extrapolation from observations at larger radii, 
we add a power-law form of the enclosed mass $M(r)\propto r^{1.6}$~\cite{GRAVITY:2020gka} in Case II.
We find that adding such an astrophysical component will make our constraints on the DM parameters slightly more stringent, 
but the impact of the astrophysical component is small, and our conclusions remain the same.

\section*{Acknowledgements} 

We thank Yi-Fu Cai, Jie-Wen Chen, Yifan Chen, Tuan Do, Kai-Kai Duan, Stefan Gillessen, Cheng-Zi Jiang, Bing Sun, Yuan-Zhu Wang, and Lei Zu for helpful discussion. This work is supported by the National Natural Science Foundation of China (NSFC) under Grants No. 12003074 and No. U1738210, and the Entrepreneurship and Innovation Program of Jiangsu Province.

\bibliographystyle{apsrev}
\bibliography{references}

\section*{Appendix}
In Table.~\ref{table1}, we report the average value and $1\sigma$ credible interval of the marginal posterior probability distribution for each parameter obtained from the MCMC calculations as well as their best-fit values in a Higgs portal model with $\mu_s=10^{-18}$~eV.
Here, we provide the 2D projections of an MCMC sample (corner plot) along with the $0.5\sigma/1.0\sigma/1.5\sigma$ contours with the {\tt corner} package~\citep{corner}.
We require all parameters satisfying the Gelman-Rubin diagnostic of $R-1<0.3$, therefore the MCMC samples are expected to be well converged.
It is worth mentioning that the mass ratio of scalar cloud $\kappa$ and Higgs portal coupling $\beta$ are converging up, because when the parameters become smaller, their contributions to S2 orbit is negligible.

\begin{figure*}[htbp]
\centering
\includegraphics[width=0.98\linewidth]{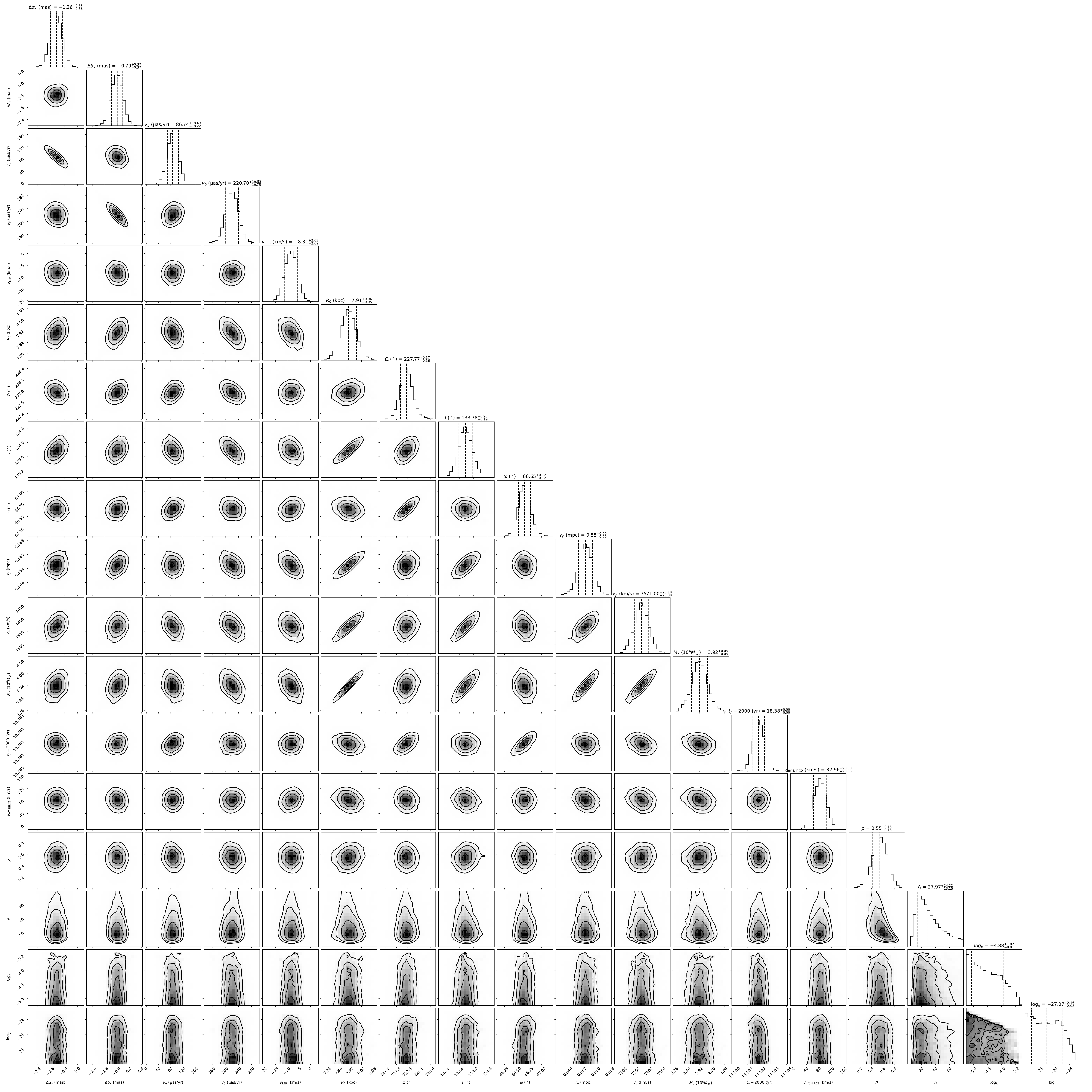}
\caption{The corner plot of 18 free parameters obtained from an MCMC sample. All parameters are well constrained, including mass ratio of scalar cloud $\kappa$ and coupling $\beta$. To ease comparisons, parameters measured in identical units are plotted with identical axes lengths.}
\label{contour}
\end{figure*}

\end{document}